# Nafion-induced reduction of manganese and its impact on the electrocatalytic properties of a highly active MnFeNi oxide for bifunctional oxygen conversion

Dulce M. Morales,*[a] Javier Villalobos,[a] Mariya A. Kazakova,[b] Jie Xiao,[c] and Marcel Risch*[a]

[a] Dr. D. M. Morales, J. Villalobos, Dr. M. Risch
Nachwuchsgruppe Gestaltung des Sauerstoffentwicklungsmechanismus
Helmholtz-Zentrum Berlin für Materialien und Energie GmbH
Hahn-Meitner-Platz 1, 14109 Berlin, Germany
E-mail: dulce.morales_hernandez@helmholtz-berlin.de, marcel.risch@helmholtz-berlin.de

[b] Dr. M. A. Kazakova
Boreskov Institute of Catalysis, SB RAS
Lavrentieva 5, 630090 Novosibirsk, Russia

[c] Dr. J. Xiao
Department of Highly Sensitive X-ray Spectroscopy
Helmholtz-Zentrum Berlin für Materialien und Energie GmbH
Albert-Einstein-Straße 15, 12489 Berlin, Germany

Supporting information for this article is given via a link at the end of the document.

**Abstract:** Electrocatalysts for bifunctional oxygen reduction (ORR) and oxygen evolution reaction (OER) are commonly studied under hydrodynamic conditions, rendering the use of binders necessary to ensure the mechanical stability of the electrode films. The presence of a binder, however, may influence the properties of the materials under examination to an unknown extent. Herein, we investigate the impact of Nafion on a highly active ORR/OER catalyst consisting of MnFeNi oxide nanoparticles supported on multi-walled carbon nanotubes. Electrochemical studies revealed that, in addition to enhancing the mechanical stability and particle connectivity, Nafion poses a major impact on the ORR selectivity, which correlates with a decrease in the valence state of Mn according to X-ray absorption spectroscopy. These findings call for awareness regarding the use of electrode additives, since in some cases the extent of their impact on the properties of electrode films cannot be regarded as negligible.

Electrocatalytic oxygen conversion–comprising the oxygen evolution (OER) and the oxygen reduction reaction (ORR)–is considered a major obstacle for the commercialization of green, regenerative energy conversion technologies, e.g., reversible fuel cells or rechargeable metal-air batteries, since both reactions suffer from sluggish kinetics.[1] Hence, the design of low-cost, high-performance, bifunctional ORR/OER electrocatalysts (BOEs) is essential to increase the efficiency of these devices.[2] Investigation of BOEs is typically conducted under hydrodynamic conditions using rotating disk electrode (RDE) setups or electrochemical flow cells.[3,4,5] It is thus necessary that the mechanical stability of the investigated electrode film is sufficiently high to endure the experimental conditions required for the evaluation of its catalytic properties. This can be achieved by means of a binder, which not only aids in preventing catalyst detachment, but also in improving the electric contact between catalyst particles and electrode substrate, lowering thus the resistance of the electrode film.[6,7] Depending on the binder and its concentration, additional effects on various properties of electrode films have been reported, including ionic conductivity, hydrophobicity, mass transport, and accessibility to the reaction sites, often leading to an improvement of the catalytic performance.[8–10]

A popular compound used as a binder is Nafion, an ion-conducting ionomer comprising hydrophilic, sulfonic-terminated side chains, and built upon copolymerization of tetra-fluoroethylene and perfluorinated vinyl ether monomers.[11] This binder has proved successful in improving the mechanical stability of electrode films, and thus it has been recommended for benchmark electrode preparation protocols.[12,13] It is reported that Nafion may induce decreases in overpotential attributed to an increased electrical conductivity and facilitated mass transport,[6] though depending largely on the electrode composition,[8,12,14] and not impacting otherwise the catalytic activity of, for instance, IrO$_2$,[10,12] Pt/C,[6,15] or Pt-Sn/C.[15] However, the nature of additional effects observed with diverse materials remains unclear. It has been speculated that Nafion impacts the intrinsic catalytic properties of Pt, attributed to the specific adsorption of sulfonate groups on the catalyst surface.[8] Moreover, it was recently reported that OER activity trend exhibited by Mn oxides of various crystal structures was different in the presence of Nafion than in its absence, speculatively due to a binder-induced chemical change of the surface of these materials.[16] Furthermore, the acidic nature of Nafion may also lead to corrosion and catalyst dissolution in the case of materials that are not chemically stable in low pH media,[5] impacting further the apparent activity of the investigated catalyst films. Hence, understanding the influence of Nafion on the catalytic properties of materials under investigation results crucial for studies that aim to elucidate their intrinsic properties.

As a case study, we investigate multiphase Mn, Ni and Fe oxide nanoparticles (10:7:3 metal ratio) supported on oxidized multi-walled carbon nanotubes, hereafter denoted MnFeNiOx, which was recently proposed as a high-performance BOE.[17] Catalyst inks (MnFeNiOx dispersed in a 1:1 water-ethanol mixture) were deposited onto glassy carbon (GC) RDEs in the presence or absence of Nafion (2 vol%) in order to observe its impact on the electrocatalytic properties of the obtained catalyst films. The binder-containing and binder-free films are hereafter denoted MnFeNiOx-Nafion/GC and MnFeNiOx/GC, respectively. Linear sweep voltammograms (LSVs) of MnFeNiOx/GC and MnFeNiOx-Nafion/GC were recorded in triplicate in the ORR and OER potential regions. The individual measurements and their averages are shown in **Figure S1** and **Figure 1a**, respectively. To facilitate the comparison, we determined the potential at which disk current densities of +10 mA cm$^{-2}$ ($E_{OER}$) and -1 mA cm$^{-2}$ ($E_{ORR}$) were attained for the two samples as activity metrics[4,18] (**Table S1**). The obtained $E_{OER}$ values were 1.547±0.006 and 1.538±0.007 V vs. RHE for MnFeNiOx/GC and MnFeNiOx-Nafion/GC, respectively, indicating only a slight increase in the apparent OER activity of the electrodes upon addition of Nafion, which could be explained by an improved contact between the conductive MnFeNiOx powder and the GC substrate.[10] Similarly in the case of the ORR, MnFeNiOx-Nafion/GC displayed an $E_{ORR}$ value of 0.788±0.003 V vs. RHE, while for MnFeNiOx/GC $E_{ORR}$ was 0.776±0.009 V vs. RHE, with a comparatively higher reproducibility in the case of the former as shown in **Figure S2**. Yet, a more substantial difference in the recorded ORR currents was observed in the kinetic-diffusion mixed control region, which could be related to differences in ORR selectivity. To investigate this, LSVs were recorded at different electrode rotation rates, and the number of electrons transferred during the ORR ($n$) was determined via the Koutecky-Levich (K-L) analysis.[19,20] The analysis was conducted on three independent sets of measurements (**Figure S2** and **S3**), displaying high reproducibility in the investigated potential range (**Table S2**). Average LSVs and K-L plots are displayed in **Figure 1b** and **1c**, respectively. For MnFeNiOx/GC, $n$ had a value of 2.3, which translates into a pathway favoring the formation of peroxide species ($n$ = 2), according to **Equation 1**.[20] Interestingly, for the Nafion-containing sample $n$ was 3.8, indicating that the direct reduction of O$_2$ to OH$^-$ is favored (**Equation 2**),[20] which agrees with an earlier study conducted by rotating ring disk electrode voltammetry.[17] These results indicate that the presence of Nafion leads to a major improvement of the ORR selectivity of the trimetallic catalyst.

$$O_2 + H_2O + 2e^- \rightarrow HO_2^- + OH^- \quad (1)$$
$$O_2 + 2H_2O + 4e^- \rightarrow 4OH^- \quad (2)$$

MnFeNiOx was initially proposed as a two-component catalyst with FeNiOx (3:7 metal ratio) being the active site for the OER,[21] and MnOx being the key component that activates the catalyst towards the ORR.[17] If this assumption is correct, and given that the presence of the binder led to substantial enhancement of the ORR performance while barely influencing the OER activity, it can be hypothesized that the catalyst undergoes Nafion-induced chemical changes related to Mn. Since correlations between Mn valence and ORR selectivity have been established,[22,23] we resorted to X-ray absorption spectroscopy (XAS) for an in-depth investigation of our hypothesis.

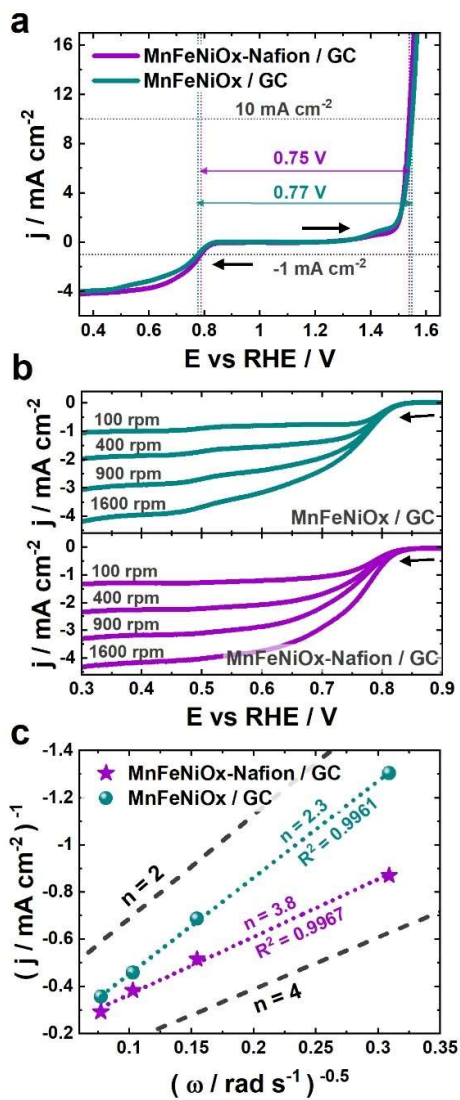

**Figure 1.** $iR_U$-drop-compensated LSVs corresponding to MnFeNiOx deposited onto GC-RDEs in the presence (purple) and in the absence (teal) of Nafion with a scan rate of 5 mV s$^{-1}$ (a) at 1600 rpm electrode rotation in the OER and ORR potential regions, (b) at different electrode rotation rates in the ORR potential region, and (c) their corresponding Koutecky-Levich plots obtained at 0.65 V vs. RHE. Simulated plots, corresponding to the transfer of 2 and 4 electrons, are shown for guidance, and were determined considering $D$ = 1.9x10$^{-5}$ cm$^2$ s$^{-1}$, $\nu$ = 1.1x10$^{-2}$ cm$^2$ s$^{-1}$, and $C$ = 1.2x10$^{-6}$ mol cm$^{-3}$.[20] All measurements were conducted in O$_2$-saturated 0.1 M NaOH solution. Black arrows indicate the direction of the voltammetric scan.

XAS spectra were collected before (MnFeNiOx) and after (MnFeNiOx(Nafion)) treating the catalyst in a Nafion-containing water-ethanol solution by sonication for 15 min (see sample preparation protocol in Supporting Information). The spectra obtained in the Ni-L$_3$, Fe-L$_3$ and Mn-L$_3$ edges are shown in **Figure 2a**, **2b** and **2c**, respectively, displaying alongside the spectra of metal oxides of unmixed oxidation states for reference. While no substantial differences in the Ni-L$_3$ spectra were shown by MnFeNiOx and MnFeNiOx(Nafion) (**Figure 2a**), in the case of the Fe-L$_3$ edge, a change in the background was observed at energies above ~712 eV (**Figure 2b**), attributed to the F-K edge absorption of Nafion's fluorine atoms (**Figure 2d**).[24] Yet, the prominent Fe-L$_3$ peak did not feature any other visible change. In the case of the Mn-L$_3$ energy region (**Figure 2c**) the spectra recorded for MnFeNiOx and MnFeNiOx(Nafion) displayed major differences. Earlier, XRD characterization revealed that MnFeNiOx consists of various mono and bimetallic oxide phases in a highly defective state.[17] Thus, Mn in MnFeNiOx is expected to be present in a mixture of oxidation states, as observed in **Figure 2c**, with a larger prevalence of Mn$^{2+}$ and Mn$^{3+}$ surface species according to XPS analyses.[17] The more intense feature centered at ~641 eV in the spectrum of MnFeNiOx(Nafion) compared to MnFeNiOx suggests that Mn became mainly present in oxidation state 2+ upon exposure to Nafion-containing solution. Given that Fe and Ni spectra did not display any evident change that correlates with a variation of their oxidation states, it can be assumed that the majority of the binder-caused surface state changes that led to an improvement of the ORR selectivity are related to Mn. This observation supports the hypothesis that Mn oxide is an essential component of the ORR active sites in MnFeNiOx.

The increase in the intensity of the spectral feature related to Mn$^{2+}$ species could be attributed to a Nafion-induced chemical reduction, which has been observed previously with Mn-containing complexes.[25] However, another plausible explanation is that the strongly acidic proton in the binder induces disproportionation of Mn$^{3+}$, forming Mn$^{2+}$ and Mn$^{4+}$,[26] and resulting in an apparent increase in Mn$^{2+}$ species due to a variation in the ratio Mn$^{2+}$:Mn$^{3+}$:Mn$^{4+}$. To further understand the origin of the change in Mn oxidation state, two control experiments were conducted with MnFeNiOx treated in (1) a water-ethanol mixture in the absence of Nafion (MnFeNiOx(WE)), and in (2) a solution containing cation-exchanged Nafion (MnFeNiOx(Nafion-Na$^+$)), namely, after having replaced the H$^+$ at the sulfonate groups with Na$^+$ according to a previously reported procedure.[5] The obtained XAS spectra are displayed in **Figure 2c**, showing that, while no substantial spectral difference was observed between MnFeNiOx(powder) and MnFeNiOx(WE), the changes observed with MnFeNiOx(Nafion) were displayed as well by MnFeNiOx(Nafion-Na$^+$), thereby suggesting that the acidic proton in the binder does not play a major role in the observed Mn valence changes. We further investigated the effect of Nafion on the XAS features of MnO$_2$ and Mn$_2$O$_3$, for which the corresponding powders were treated in a water-ethanol mixture in the absence (Mn$_X$O$_Y$(WE)) or presence (Mn$_X$O$_Y$(Nafion)) of the binder (**Figure 2e**). Interestingly, the peak assigned to Mn$^{2+}$ (641 eV) was clearly seen with MnO$_2$ (Mn$^{4+}$) upon exposure to Nafion, whereas the spectra corresponding to Mn$_2$O$_3$ (Mn$^{3+}$) did not display any substantial change, indicating that the valence changes are not related to disproportionation. We speculate that a strong chemical interaction between Mn$^{4+}$ species in MnO$_2$ and the electron donors in the binder takes place. Likely, this occurs similarly with Mn$^{4+}$ species in MnFeNiOx, thus leading to an overall improvement in the ORR performance of the catalyst: on the one hand, the formation of Mn$^{2+}$ species (from Mn$^{4+}$) could favor the binding of *OOH intermediates, according to recent DFT predictions,[27] and on the other hand, the unaffected Mn$^{3+}$ atoms provide O$_2$ absorption sites[28] and a Mn$^{3+}$:Mn$^{4+}$ ratio that facilitates the 4-electron transfer pathway.[23,29]

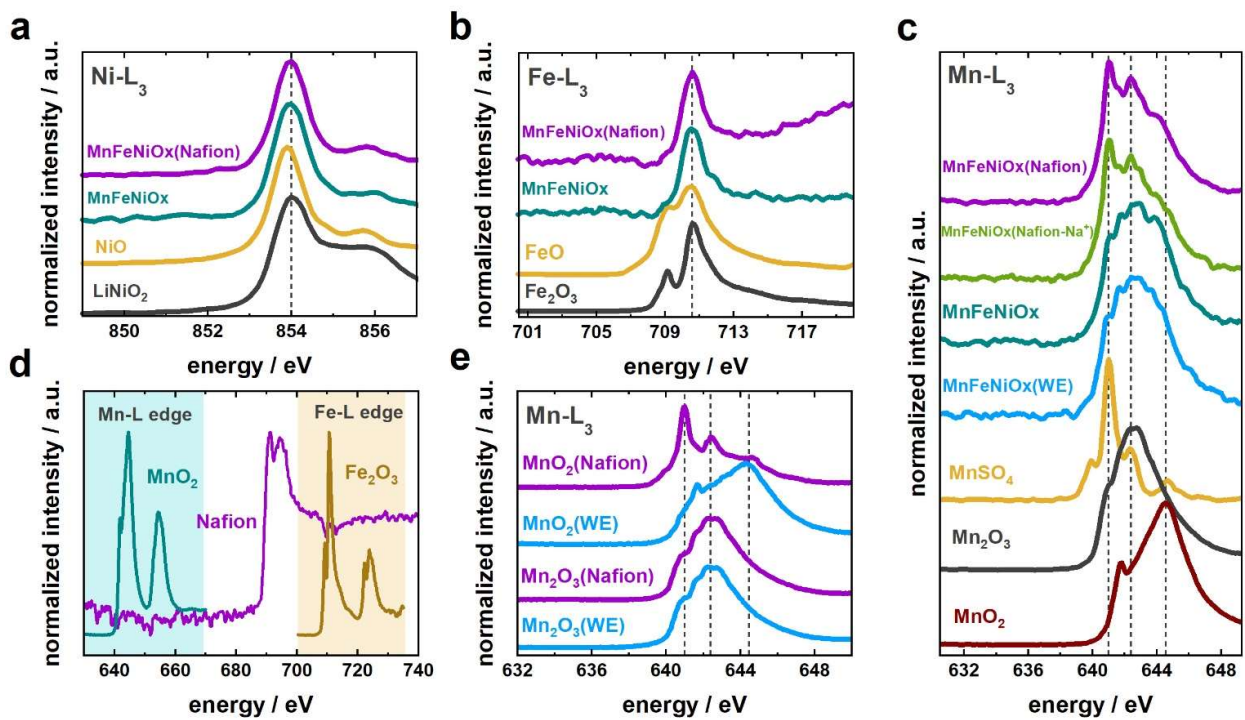

**Figure 2.** Normalized XAS spectra of MnFeNiOx recorded in total electron yield mode before and after treatment in water-ethanol mixtures without (WE) or with 2 vol% binder (Nafion), recorded in the L$_3$ edge of (a) Ni, (b) Fe, and (c) Mn, showing corresponding reference compounds. (c) shows additionally the spectra of MnFeNiOx exposed to a WE solution containing cation-exchanged binder (Nafion(Na$^+$)). (d) XAS spectrum of Nafion in the energy region comprising both the L$_2$-L$_3$ edge of Mn and the L$_2$-L$_3$ edge of Fe; the spectra of MnO$_2$ and Fe$_2$O$_3$ are shown for reference. Spectra were offset for clarity and vertical dashed lines are included to guide the eye.

In summary, we investigated the influence of Nafion on the electrocatalytic properties of MnFeNiOx as a bifunctional ORR/OER catalyst. Besides the advantageous, yet expected, improvement of electrode film properties (mechanical stability and particle contact), a major benefit on the ORR selectivity of MnFeNiOx was revealed: while binder-free MnFeNiOx displayed a preferred 2-electron transfer pathway, in the presence of Nafion the catalyst rather exhibited the direct reduction of $O_2$ to $OH^-$ via the transfer of 4 electrons. The impressive improvement in selectivity is attributed to a binder-induced decrease in the oxidation state of Mn, as observed during XAS investigations, resulting in a more favorable $Mn^{2+}:Mn^{3+}:Mn^{4+}$ ratio, and confirming that Mn plays a major role in the ORR performance of the trimetallic catalyst. Control experiments on MnFeNiOx and commercial Mn oxides indicated, on the one hand, that the valence changes observed are neither related to the acidic nature of Nafion, nor due to disproportionation, and on the other hand, that $Mn^{4+}$ species are susceptible to chemical reduction in the presence of Nafion. Although further studies are still required to fully reveal the extent of the chemical changes, as well as the variety of materials that may be susceptible to them, it is clear that Nafion cannot always be regarded as inert, and that awareness on its use is required for investigations where the intrinsic activity of a catalytic material is the main focus of the work.

## Acknowledgments

The authors are grateful to Dr. Ronny Golnak, Denis Antipin, Joaquín Morales-Santelices, and Sepideh Madadkhani for their support during collection of XAS data, and to Christian Höhn (HZB, Institute for Solar Fuels) for preliminary XPS analyses. We thank HZB for the allocation of synchrotron radiation beamtime. This project has received funding from the European Research Council (ERC) under the European Union's Horizon 2020 research and innovation programme under grant agreement No 804092. M. A. K. acknowledges the support of the Ministry of Science and Higher Education of the Russian Federation within the state assignment for Boreskov Institute of Catalysis (project # AAAA-A21-121011390054-1).

**Keywords:** bifunctional oxygen electrodes • Nafion • oxidation state • transition metals • X-ray absorption spectroscopy

# Supporting Information

# Nafion-induced reduction of manganese and its impact on the electrocatalytic properties of a highly active MnFeNi oxide for bifunctional oxygen conversion

Dulce M. Morales,*[a] Javier Villalobos,[a] Mariya A. Kazakova,[b] Jie Xiao,[c] and Marcel Risch*[a]

[a]  Nachwuchsgruppe Gestaltung des Sauerstoffentwicklungsmechanismus, Helmholtz-Zentrum Berlin für Materialien und Energie GmbH, Hahn-Meitner-Platz 1, 14109 Berlin, Germany

[b]  Boreskov Institute of Catalysis, SB RAS, Lavrentieva 5, 630090 Novosibirsk, Russia

[c]  Department of Highly Sensitive X-ray Spectroscopy, Helmholtz-Zentrum Berlin für Materialien und Energie GmbH, Hahn-Meitner-Platz 1, 14109 Berlin, Germany

*Corresponding author e-mail:

dulce.morales_hernandez@helmholtz-berlin.de (Dulce M. Morales)

marcel.risch@helmholtz-berlin.de (Marcel Risch)

## 1. Experimental

*Chemicals*

Powdered $MnSO_4$, $Mn_2O_3$, $MnO_2$, $LiNiO_2$ and FeO, as well as Nafion solution (5 vol%) were purchased from Sigma-Aldrich. NiO, and $Fe_2O_3$ were purchased from Roth and Alfa Aesar, respectively. These chemicals were used as received without further purification.

Synthesis of MnFeNiOx catalyst was reported previously.[1] Firstly, growth of multi-walled carbon nanotubes (MWCNTs) was carried out by chemical vapor deposition of ethylene at 680 °C using $Fe_2Co/Al_2O_3$ as catalyst, with subsequent acidic treatment in HCl for 4 h to remove catalyst residues. After washing to neutral pH and drying, oxygen functionalities were introduced to the obtained powder by boiling it in concentrated $HNO_3$ for 2 h, followed by washing with distilled water and drying, thus obtaining oxidized MWCNTs (MWCNTs-Ox). A solution containing a mixture of Mn(II), Ni(II) and Fe(III) nitrates (Sigma-Aldrich) with metal ratio of 10:7:3 was used for preparation of MnFeNiOx/MWCNTs-Ox catalyst via incipient wetness impregnation. The obtained material was dried at 110 °C for 4 h, and subsequently annealed at 350 °C for 4 h in inert atmosphere, forming thus trimetallic oxide nanoparticles with a total metal loading of 14.4 wt%, and individual metal loadings of 7.3, 5.0, and 2.1 wt% corresponding to Mn, Ni, and Fe, respectively, according to XRF studies.[1]

*Electrochemical methods*

Electrochemical measurements were conducted following the procedure illustrated in **Scheme S1**, and described in detail in this section.

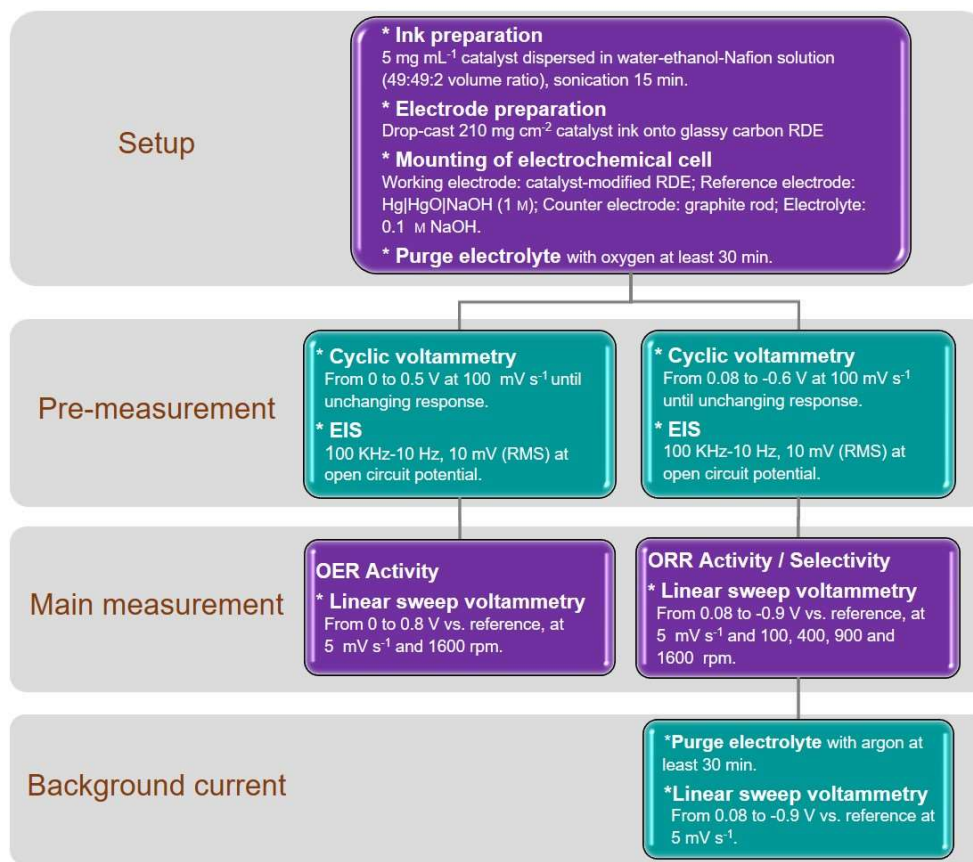

**Scheme S1.** Electrochemical protocol used for the investigation of catalytic properties of MnFeNiOx.

Electrochemical investigations were conducted in a three-electrode configuration, single-compartment electrochemical cell. The setup was kept into a climate chamber (Binder) set to 25 °C during the measurements to ensure temperature control. The working, counter and reference electrodes were a catalyst-coated glassy carbon (GC) rotating disk electrode (4 mm diameter), a graphite rod, and a Hg|HgO|NaOH (1 M) electrode (ALS Inc.), respectively. Catalyst deposition onto the GC electrode was done by drop-casting 5.3 µL catalyst ink, which consisted of 5 mg mL$^{-1}$ dispersion of the active material in a mixture of water and ethanol (1:1 volume ratio), in the presence or absence of 2 vol% Nafion solution, by sonication for 15 min. The total catalyst loading on the GC electrode was 210 µg cm$^{-2}$. The electrolyte was 0.1 M NaOH standard solution (Sigma-Aldrich) thoroughly purged with oxygen or argon. A flowing stream of the corresponding gas was kept onto the electrolyte surface during the measurements to maintain the gas saturation. All measurements were conducted using a Reference 600+ potentiostat (Gamry) equipped with an RRDE-3A rotator (ALS Inc.).

Prior to the measurements, catalyst-modified electrodes were subjected to continuous potential cycling at a scan rate of 100 mV s$^{-1}$ in the potential range from 0.08 to -0.6 V vs. Hg|HgO|NaOH (prior to ORR) or from 0 to 0.5 V vs. Hg|HgO|NaOH (prior to OER) until an unchanging voltammetric response was observed. Electrochemical impedance spectra were subsequently recorded in the frequency range from 100 kHz to 1 Hz with an AC amplitude of 10 mV (RMS).

The uncompensated resistance ($R_U$) was determined from the resulting Nyquist plots, and later used to $iR_U$-drop-correct the measured potentials according to **Equation 1**, where $i$ is the measured current. The obtained $R_U$ values were in average 47 ± 4 Ω.

$$E_{corrected} = E_{measured} - i\, R_U \qquad (1)$$

Linear sweep voltammograms were afterwards recorded in the potential ranges from 0.08 to -0.9 V vs. Hg|HgO|NaOH and from 0 to 0.8 V vs. Hg|HgO|NaOH, for the ORR and for the OER, respectively, at a scan rate of 5 mV s$^{-1}$ and rotation rate of 1600 rpm to evaluate the catalytic activity towards the corresponding reaction. An additional voltammogram was recorded in Ar-purged electrolyte at the same scan rate to determine the background current. All measurements were done at least in triplicate. The individual background-corrected measurements used for obtaining the average voltammograms depicted in **Figure 1a** in the main manuscript are shown in **Figure S1**.

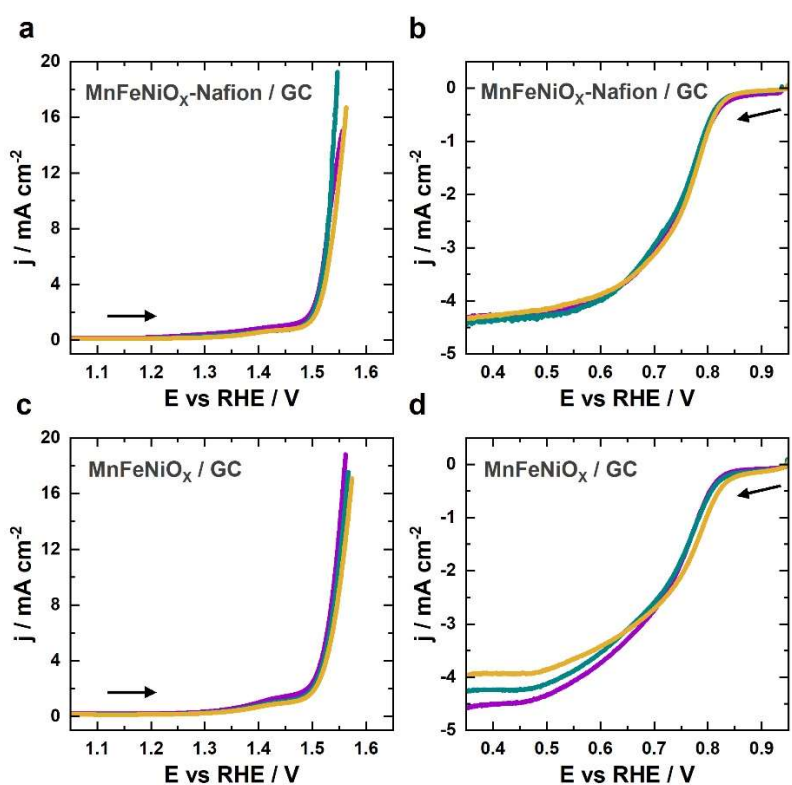

**Figure S1.** Linear sweep voltammograms of MnFeNiOx recorded in triplicate in O$_2$-saturated 0.1 M NaOH solution at a scan rate of 5 mV s$^{-1}$ and electrode rotation of 1600 rpm in the (a,c) OER, and (b,d) ORR potential regions. Each voltammogram was recorded with a freshly prepared electrode film (a,b) in the presence and (c,d) in the absence of Nafion. Black arrows indicate the direction of the voltammetric scan.

All potentials are reported with respect to the reversible hydrogen electrode (RHE) scale. To convert the measured potentials to this scale, at the start of each experiment day the voltage between the reference electrode and an RHE electrode (Gaskatel) was measured for 10 min. The last value recorded was registered and added to the potentials measured. The average value obtained for different experiment days was 0.8795 ± 0.0216 V. Activity metrics $E_{OER}$ and $E_{ORR}$, corresponding to the potentials vs. RHE at which current densities of +10 and -1 mA cm$^{-2}$ were attained, respectively, were determined from the iR$_U$-compensated voltammograms, and the obtained values are shown in **Table S1**. These activity metrics were chosen according to reported guidelines.[2,3]

**Table S1.** Activity metrics corresponding.

| Sample | Activity metric | E vs. RHE / V set 1 | E vs. RHE / V set 2 | E vs. RHE / V set 3 | E vs. RHE / V average |
|---|---|---|---|---|---|
| MnFeNiOx-Nafion/GC | $E_{OER}$[a] | 1.535 | 1.533 | 1.546 | 1.538 ± 0.007 |
|  | $E_{ORR}$[b] | 0.789 | 0.785 | 0.791 | 0.788 ± 0.003 |
| MnFeNiOx/GC | $E_{OER}$[a] | 1.542 | 1.547 | 1.553 | 1.547 ± 0.006 |
|  | $E_{ORR}$[b] | 0.785 | 0.772 | 0.771 | 0.776 ± 0.009 |

[a] E vs RHE at +10 mA cm$^{-2}$; [b] E vs RHE at -1 mA cm$^{-2}$.

ORR selectivity was investigated by collecting linear sweep voltammograms in the ORR potential region at rotation rates of 100, 400, 900 and 1600 rpm. The background-corrected voltammograms reported in the main manuscript represent the average of the three independent sets of measurements shown in **Figure S2**.

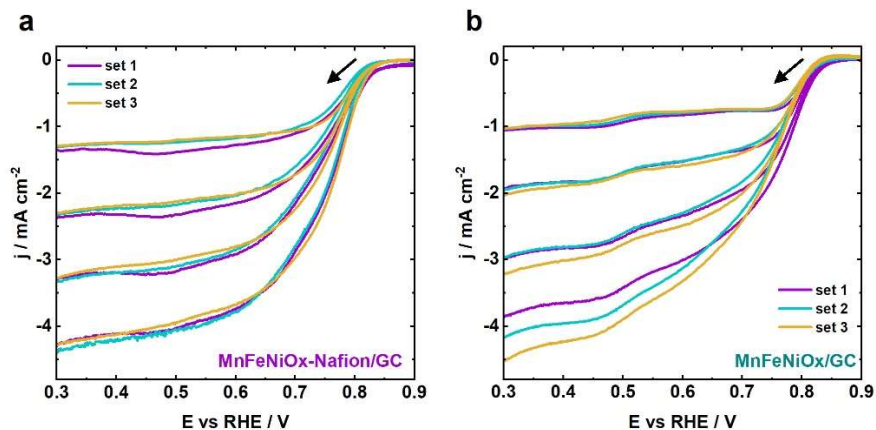

**Figure S2.** Comparison of three independent sets of background-corrected linear sweep voltammograms recorded at a scan rate of 5 mV s$^{-1}$ and electrode rotation rates of 100, 400, 900 and 1600 rpm in O$_2$-saturated 0.1 M NaOH solution, corresponding to (a) MnFeNiOx-Nafion/GC, and (b) MnFeNiOx/GC. Black arrows indicate the direction of the voltammetric scan.

Subsequently, current density (*j*) was extracted from the background-corrected voltammograms at selected potentials, and the inverse of the obtained values was later plotted as a function of the square root of the angular velocity of rotation ($\omega$), which is related to the electrode rotation rate (*r*) according to **Equation 2**. The linear regression obtained from the plotted data is described by the Koutecky-Levich equation (**Equation 3**):[3,4]

$$\omega = \frac{2\,r\,\pi}{60} \qquad (2)$$

$$\frac{1}{j} = \frac{1}{j_k} + \frac{1}{0.62\,n\,F\,D^{2/3}\,\nu^{-1/6}\,C} \cdot \frac{1}{\omega^{1/2}} \qquad (3)$$

where, $j_k$ is the kinetic-limited current, $F$ is the Faraday constant, $D$ is the diffusion coefficient, $\nu$ is the kinematic viscosity, $C$ is the bulk concentration, and $n$ is the number of electrons transferred. With the resulting slope and considering values of $D = 1.9 \times 10^{-5}$ cm$^2$ s$^{-1}$, $\nu = 1.1 \times 10^{-2}$ cm$^2$ s$^{-1}$, and $C = 1.2 \times 10^{-6}$ mol cm$^{-3}$, corresponding to an O$_2$-saturated 0.1 M NaOH solution,[5] $n$ was determined. **Figure S3** shows the Koutecky-Levich plots obtained with three independent measurement sets. The corresponding $n$ values obtained at different potentials are summarized in **Table S2**.

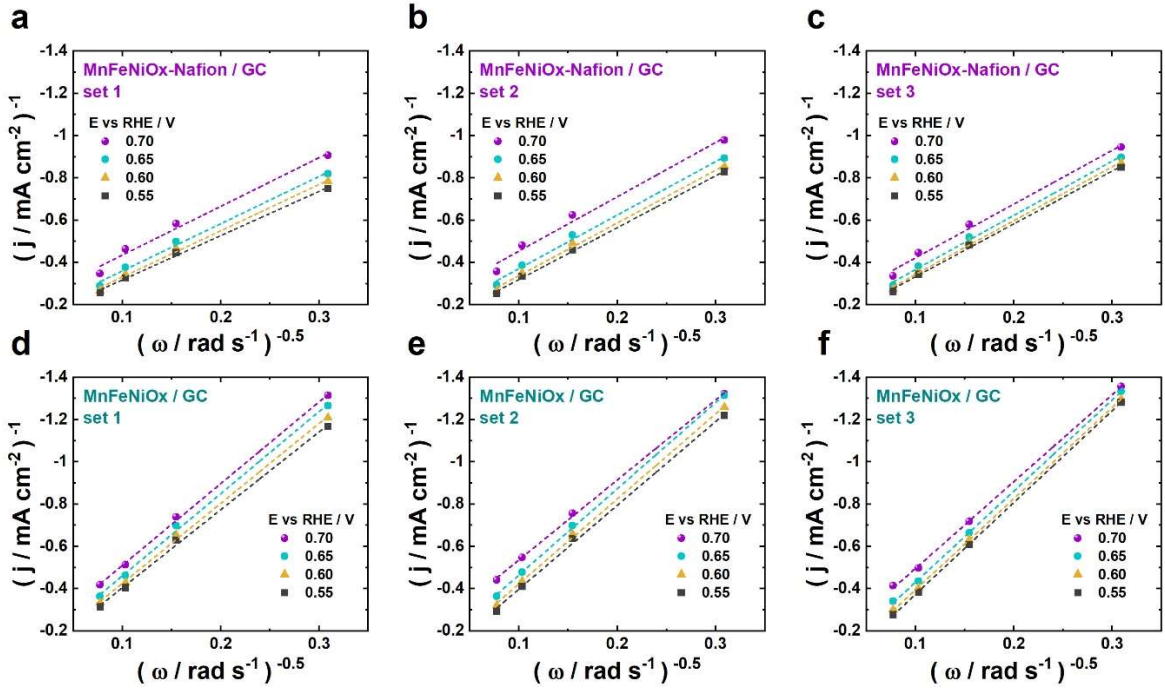

**Figure S3.** Koutecky-Levich plots obtained from three independent sets of linear sweep voltammograms recorded at a scan rate of 5 mV s$^{-1}$ and electrode rotation rates of 100, 400, 900 and 1600 rpm in O$_2$-saturated 0.1 M NaOH solution, corresponding to (a,b,c) MnFeNiOx-Nafion/GC, and (d,e,f) MnFeNiOx/GC.

**Table S2.** Number of electrons transferred (n) determined from three individual sets of experiments.

| Sample | E vs. RHE / V | n / - set 1 | n / - set 2 | n / - set 3 | n / - average |
|---|---|---|---|---|---|
| MnFeNiOx-Nafion/GC | 0.70 | 4.00 | 3.59 | 3.63 | 3.74 ± 0.226 |
|  | 0.65 | 4.13 | 3.64 | 3.60 | 3.79 ± 0.295 |
|  | 0.60 | 4.21 | 3.70 | 3.62 | 3.84 ± 0.320 |
|  | 0.55 | 4.41 | 3.76 | 3.69 | 3.95 ± 0.397 |
| MnFeNiOx/GC | 0.70 | 2.39 | 2.44 | 2.25 | 2.36 ± 0.098 |
|  | 0.65 | 2.34 | 2.26 | 2.14 | 2.25 ± 0.101 |
|  | 0.60 | 2.44 | 2.30 | 2.14 | 2.29 ± 0.150 |
|  | 0.55 | 2.50 | 2.33 | 2.12 | 2.32 ± 0.190 |

*X-ray absorption spectroscopy (XAS)*

Sample preparation was done either by attaching untreated, powdered samples to carbon tape, or by dispersing the powder (5 mg mL$^{-1}$) in one of the following dispersion solutions by sonication for 15 min:

1. Binder-free solution (WE): a mixture of water and ethanol (1:1 volume ratio).
2. Solution containing untreated binder (Nafion): a mixture of water and ethanol (1:1 volume ratio) containing 2 vol%. Nafion solution.
3. Solution containing ion-exchanged Nafion (Nafion-Na$^+$): a mixture of water and ethanol (1:1 volume ratio) with 2 vol%. Nafion solution treated according to a procedure reported elsewhere,[6] consisting of drop-wise mixing Nafion solution and 0.1 M NaOH solution (2:1 volume ratio), thus exchanging H$^+$ for Na$^+$ ions.

Binder-containing dispersions were drop-cast onto glassy carbon plates or graphite foil of 5x5 mm$^2$ and left to dry at ambient conditions. Binder-free dispersions were poured onto a watch glass and left to dry at ambient conditions. The recovered powders were pressed onto carbon tape.

XAS measurements were carried out at the LiXEdrom experimental station at the U49/2 PGM-1 beamline at the BESSY II synchrotron (Helmholtz-Zentrum Berlin für Materialien und Energie) at room temperature.[7] The samples on carbon supports were attached to a current collector using Cu tape, and the spectra were recorded in total electron yield mode by collecting the drain current with a Keithley 6514 ammeter. Spectra were collected in the ranges from 630 to 676 eV, from 700 to 735 eV, and from 832 to 885 eV, corresponding to the Mn-L$_3$, Fe-L$_3$ and Ni-L$_3$ edge energy regions, respectively, at least in duplicate and in different spots on the samples to ensure reproducibility of spectra as well as to prevent radiation-induced sample damage. Calibration of energy axis and data processing were done as reported previously,[8] using the software Bessy. In short, the recorded energies were corrected by the difference in the position of the peak of maximum intensity in the Mn-L$_3$ edge spectrum of MnSO$_4$ with respect to 641 eV. Normalization of the spectra was conducted by dividing the recorded intensities by the photon flux, followed by subtraction of the polynomial fit (order 0 or 1) of the signal before the L$_3$ edge, and subsequent division by the polynomial fit (order 0 or 1) of the signals recorded after the L$_2$ edge. A step-by-step example can be found in the supporting information of a report by Villalobos, *et al*.[8] The spectra were further normalized to a maximum intensity of 1 to facilitate their comparison.